\documentclass[cite]{PoS}

\newcommand{\nc}[1]{\newcommand{#1}}
\nc{\ben}{\begin{enumerate}}
\nc{\een}{\end{enumerate}}

\nc{\bo}[1]{\mbox{\boldmath \( #1 \! \! \)  \unboldmath}}

\nc{\be}{\begin{eqnarray}}
\nc{\ee}{\end{eqnarray}}
\nc{\la}{\langle}
\nc{\ra}{\rangle}
\def\simge{\mathrel{%
       \rlap{\raise 0.511ex \hbox{$>$}}{\lower 0.511ex \hbox{$\sim$}}}}
\def\simle{\mathrel{
       \rlap{\raise 0.511ex \hbox{$<$}}{\lower 0.511ex \hbox{$\sim$}}}}
\nc{\Fsi}{F^{\bf 1}}
\nc{\Fad}{F^{\bf 8}}
\nc{\Fsy}{F^{\bf 6}}
\nc{\Fan}{F^{{\bf 3}^*}}
\nc{\Fm}{F^M}
\nc{\tr}{{\rm tr \,}}
\nc{\Tr}{{\rm Tr \,}}
\nc{\bx}{{\bf x}}
\nc{\by}{{\bf y}}
\nc{\bz}{{\bf 0}}
\title{Heavy-quark free energy at finite temperature with 2+1 flavors of improved
Wilson quarks in fixed scale approach}

\ShortTitle{Heavy-quark free energy at finite temperature with 2+1 flavors of
improved Wilson quarks}

\author{\speaker{Yu Maezawa}$^{\, \, a}$, S.~Aoki$^b$, S.~Ejiri$^c$, T.~Hatsuda$^d$, K.~Kanaya$^b$, H.~Ohno$^b$ and T.~Umeda$^e$ (WHOT-QCD Collaboration)\\
$^a$En'yo Radiation Laboratory, Nishina Accelerator Research Center, RIKEN, Wako, Saitama 351-0198, Japan \\
$^b$Graduate School of Pure and Applied Sciences, University of Tsukuba, Tsukuba, Ibaraki 305-8571, Japan\\
$^c$Physics Department, Brookhaven National Laboratory, Upton, New York 11973, USA\\
$^d$Department of Physics, The University of Tokyo, Tokyo 113-0033, Japan\\
$^e$Graduate School of Education, Hiroshima University, Hiroshima 739-8524, Japan\\
E-mail: \email{maezawa@ribf.riken.jp}}

\abstract{
The free energy between a static quark and an antiquark is 
 studied by using the color-singlet Polyakov-line correlation 
 at finite temperature.
We perform simulations on $32^3 \times 12$, 10, 8, 6, 4 lattices
  in the high temperature phase with the RG-improved gluon action and
   2+1 flavors of the clover-improved Wilson quark action. 
Since the simulations are based on the fixed scale approach that 
 the temperature can be varied without changing the spatial volume and renormalization factor,
  it is possible to investigate temperature dependence of the heavy-quark free energy
   without any adjustment of the overall constant.
We find that, the heavy-quark free energies at short distance converge
  to the  heavy-quark potential evaluated from the Wilson-loop operator at zero temperature, 
  in accordance with the expected insensitivity of short distance physics to the temperature.
At long distance, the heavy-quark free energies approach to 
 twice the single-quark free energies, implying that the interaction between heavy quarks
 is screened.
 The Debye screening mass obtained from the long range behavior of the heavy-quark free energy
 is compared with results of the thermal perturbation theory 
  and those of $N_f=2$ and $N_f=0$ lattice simulations.
}

\FullConference{The XXVII International Symposium on Lattice Field Theory\\
                 July 26-31, 2009\\
                 Peking University, Beijing, China}

\begin{document}

%--------------------------------------------------------------------
\section{Introduction}\label{intro}

An interaction between a heavy quark and an antiquark in hot QCD medium 
 is one of the most important quantities 
  to characterize internal properties of the quark-gluon plasma (QGP).
Experimentally, it is
 related to the fate of the charmoniums and bottomoniums
  in QGP created in relativistic heavy ion collisions.
The interaction between heavy quarks at finite temperature ($T$) 
have been studied via heavy-quark free energies
  in the quenched approximation \cite{Kaczmarek:1999mm,Nakamura1,Nakamura2}, 
 in $N_f=2$ QCD with the staggered \cite{Kaczmarek:2005ui} and 
  the Wilson quark actions \cite{Bornyakov:2004ii,WHOT-M}, and 
 in $N_f = 2+1$ QCD with the staggered quark action \cite{Fodor:2007mi,Petrov:2007ug}.
In this report, we present resent studies on the heavy-quark free energy 
 in dynamical simulations of $N_f =2+1$ QCD with the Wilson quark action.
We perform finite-temperature simulations \cite{Kanaya} on $32^3 \times(12$--$4)$ lattices adopting 
 the fixed scale approach 
 which enables direct comparison of the heavy-quark free energy at different $T$
 since the spatial volume and the renormalization factor are common to any $T$ \cite{WHOT-U}.
The corresponding zero-temperature configurations 
 are taken from the results of the CP-PACS and JLQCD Collaborations \cite{CP_JL}.
The free energies between heavy quarks at finite $T$ are evaluated from the Polyakov-line correlator
 and compared with a heavy-quark potential at $T=0$ calculated from the Wilson loop operator.
We also extract the Debye screening mass, $m_D(T)$,
 by fitting the heavy-quark free energy with the screened Coulomb form,
  and compare it with results of thermal perturbation theory
   and of lattice simulations in $N_f=2$ and $N_f=0$ QCD.

It is found that the heavy-quark free energy is insensitive to $T$ at short distance region,
 whereas it is screened by thermal medium at long distance region and
  converges to twice a single-quark free energy.
The Debye screening mass shows better agreement with that calculated in the next-to-leading order 
  of thermal perturbation than in the leading order.
Also it receives a sizable effects from the dynamical light quarks.

\section{Heavy-quark free energy}
At $T=0$, interaction between a static quark and an antiquark 
 can be studied by the heavy-quark potential evaluated from the Wilson loop operator.
 The resulting potential takes the Coulomb form at short distances 
 due to perturbative gluon exchange, while it takes 
 the linear form  at long distances due to confinement:
\be
 V(r) = - \frac{\alpha_0}{r} + \sigma r + V_0 
.
\label{eq:V}
\ee
For $T>0$, inter-quark interaction may be studied by the heavy-quark free energy
 $\Fsi (r,T)$
 evaluated from a Polyakov-line  correlation function in the color-singlet channel 
  with Coulomb gauge fixing \cite{Nadkarni}:
\be
\Fsi (r,T) = - T \ln \la \Tr \Omega^\dagger (\bx) \Omega (\by) \ra
,
\ee
where $r=|\bx - \by|$ and $\Omega(\bx) = \prod_{\tau=1}^{N_t} U_4(\bx,\tau)$
with $U_4(\bx,\tau)$ being the link variable in the temporal direction. 
At zero temperature, we expect $\Fsi(r,T=0) = V(r)$, while
at high temperature, $\Fsi(r,T)$ is screened by thermal medium and may behave as
 the screened Coulomb form:
\be
\Fsi(r,T) = - \frac{\alpha(T)}{r} e^{-m_D(T)\, r}
,
\label{eq:mD}
\ee
where $\alpha$ and $m_D$ are the effective coupling and the Debye screening mass, respectively.

\section{Fixed scale approach}
 In the conventional fixed $N_t$ approach where $T$ is varied by changing the lattice spacing $a$,  
 $V(r)$ and $\Fsi(r,T)$ receive different 
renormalization at each $T$, and thus are usually adjusted by hand such that
 $V(r)$ and $\Fsi(r,T)$ coincide with each other at a short distance,
  assuming that the short distance properties are insensitive to the temperature. 
On the other hand, in the fixed scale approach, 
the temperature $T=(N_t a)^{-1}$ is varied by changing the temporal 
lattice size $N_t$ at fixed $a$ \cite{WHOT-U}.

In the fixed scale approach, because the coupling parameters are the same for all temperatures,
 the renormalization factors are common to all temperatures.
The spatial volume is also the same.
We can thus directly compare the free energies at different temperatures
 without any adjustment. 
We show below that $\Fsi(r,T)$ for different $T$ approaches to $V(r)$ at short distances, 
 which proves the expected insensitivity of the short distance physics to the temperature.

%--------------------------------------------------------------------
\section{Results of the lattice simulations}

We employ the renormalization-group improved gluon action and $2+1$ flavors of nonperturbatively 
 $O(a)$-improved Wilson quark actions.
 Zero-temperature configurations are given by the CP-PACS and JLQCD Collaborations \cite{CP_JL}
 at $m_{\pi}/m_{\rho}=0.6337(38)$ and $m_K/m_{K^\ast} = 0.7377(28)$.  
 Finite temperature simulations with the same parameters are performed 
  on $32^3 \times N_t$ lattices with $N_t=12$, 10, 8, 6 and 4, which 
  correspond to $T \sim 200$--700 MeV \cite{Kanaya}.
We generate full QCD configurations by the hybrid Monte Carlo algorithm
 and measure the heavy-quark free energy using 500--700 configurations 
  at every 5 trajectories after thermalization.
The statistical errors are determined 
 by the jackknife method with the bin-size of 20 trajectories.
 The absolute scale is estimated from the Sommer parameter, $r_0=0.5$ fm.

\begin{figure}[bt]
  \begin{center}
    \includegraphics[width=140mm]{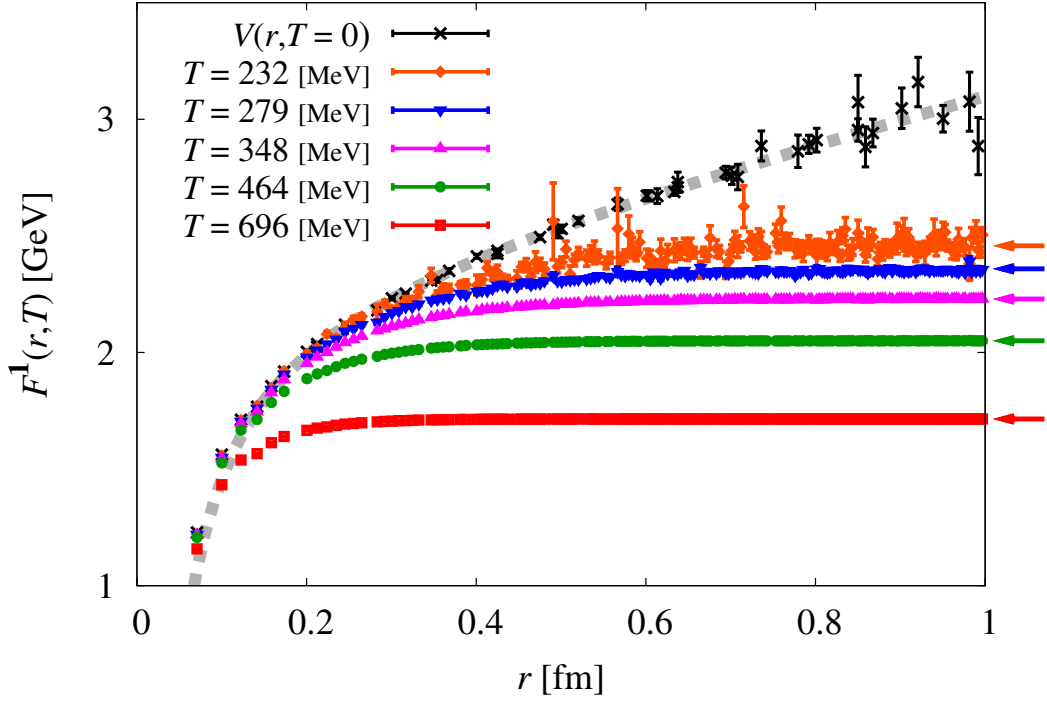}
    \caption{Heavy-quark free energies at various temperatures. 
    The heavy-quark potential at $T=0$ was calculated by the CP-PACS and JLQCD Collaborations \cite{CP_JL}.
    The fit result of $V(r)$ by the Coulomb $+$ linear form is shown by the dashed line.
    The arrows on the right side denote twice the thermal average 
    of the single-quark free energy defined as $ 2 F_Q = - 2 T \ln \la {\rm Tr} \Omega \ra $.
    }
    \label{fig:1}
  \end{center}
\end{figure}

Figure \ref{fig:1} shows the results of the heavy-quark potential $V(r)$ at $T=0$ and
 the heavy-quark free energies $\Fsi(r,T)$ at various temperatures as functions of $r$.
At $T=0$, $V(r)$ shows Coulomb-like and linear-like behaviors at short and long distances, respectively.
A fit with Eq.~(\ref{eq:V}), as shown by the dashed line in Fig.~\ref{fig:1}, gives $\alpha_0 = 0.441$ and $\sqrt{\sigma} = 0.434$ GeV.

For $T>0$, we note that the heavy-quark free energies $\Fsi(r,T)$ at all 
temperatures converge to $V(r)$ at short distances.
This means that the short distance physics is insensitive to temperature.
As stressed above, unlike the case of the conventional fixed-$N_t$ approach
in which this insensitivity is assumed and  used to adjust the constant terms of $\Fsi(r,T)$,
 our fixed scale approach enabled us to directly confirm this theoretical expectation.

At large $r$, $\Fsi(r,T)$ departs from $V(r)$
 and eventually becomes flat due to  Debye screening.
  In  Fig.~\ref{fig:1}, the asymptotic values of $\Fsi(r,T)$ at long distance 
  are also compared with $2F_Q$ denoted by the arrows where the thermal average
  of a single Polyakov-line is defined as $F_Q=-T \ln \la \Tr \Omega \ra$.
  We find that $\Fsi(r,T)$ converges to $2F_Q$ quite accurately at long distances.

\section{Debye screening mass}

\begin{figure}[bt]
  \begin{center}
    \begin{tabular}{cc}
    \includegraphics[width=72mm]{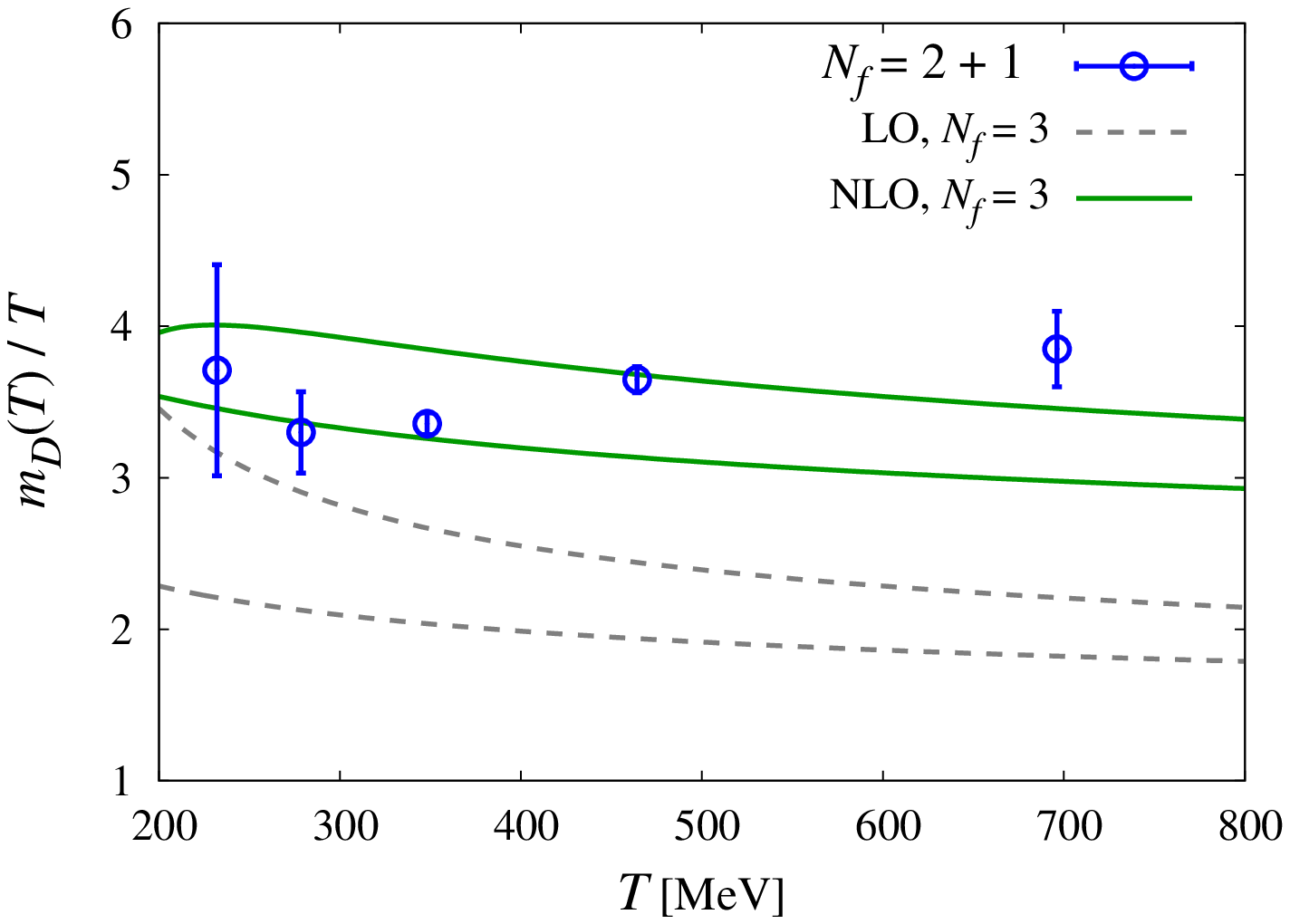} &
    \includegraphics[width=72mm]{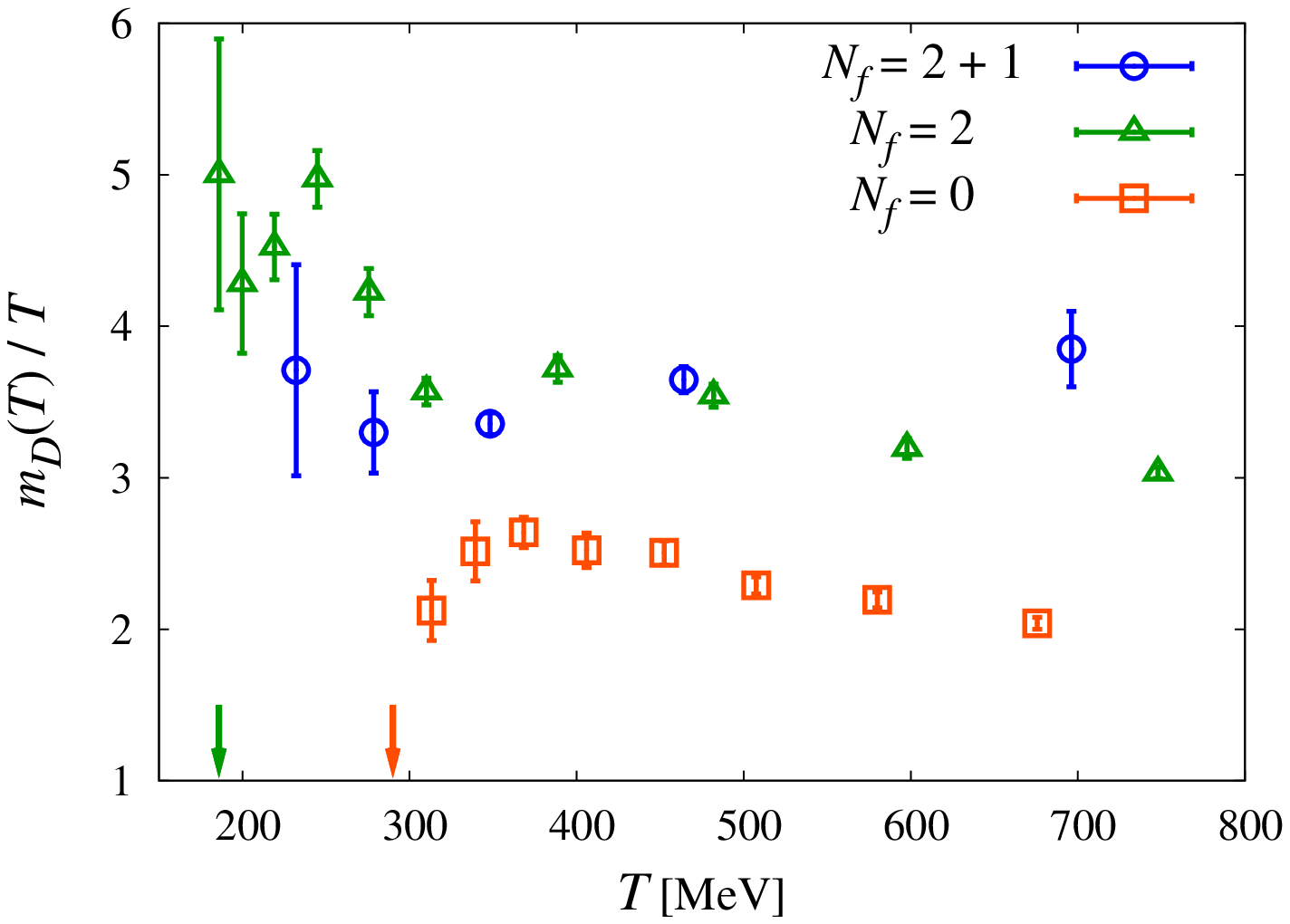}
    \end{tabular}
    \caption{Left: Results of the Debye mass for $N_f=2+1$ obtained from the heavy-quark
     free energy with
   those of the leading order (LO) and next-to-leading order (NLO) thermal perturbation theory 
   for $N_f=3$. Right: Comparison among the Debye masses for $N_f=2+1$, $N_f=2$ \cite{WHOT-M}
   and $N_f=0$ \cite{WHOT-U}
   extracted from the heavy-quark free energy on the lattice.
   Arrows in the right panel indicate critical temperatures in $N_f=2$ and $N_f=0$ cases.}
    \label{fig:2}
  \end{center}
\end{figure}

In order to extract the screening mass $m_D$,
 we fit $\Fsi(r,T) - 2F_Q$ by the screened Coulomb form, Eq.~(\ref{eq:mD}).
To determine the appropriate fit range, we estimate the effective Debye mass
 from the ratio of $\Fsi(r,T)$:
\be
m_D^{\rm eff} (T;r) = \frac{1}{\Delta r} \log \frac{\Fsi(r)-2F_Q}{\Fsi(r+\Delta r)-2F_Q}
                     - \frac{1}{\Delta r} \log \left[ 1 + \frac{\Delta r}{r} \right]
.
\ee
Investigating the plateau of $m_D^{\rm eff}(T;r)$, we choose the fit range to be
 0.35 fm $\simle r \simle$ 0.56 fm.
Results of $m_D(T)/T$ are shown in the left panel of Fig.~\ref{fig:2}, and
 does not have strong dependence on $T$.
This property is qualitatively consistent with the prediction in thermal perturbation theory:
 $m_D \sim gT$ where $g$ is the running coupling.

To make a  quantitative comparison to the result of the thermal perturbation theory,
 we define the  2-loop running coupling by
\be
g^{-2}_{\rm 2l} (\mu) 
   =  \beta_0 \ln ( \frac{\mu}{\Lambda_{\overline{\rm MS}}} )^2 + 
\frac{\beta_1}{\beta_0} 
\ln \left[ \ln ( \frac{\mu}{\Lambda_{\overline{\rm MS}}} )^2 \right]
\ee
with the QCD scale parameter 
 $\Lambda_{\overline{\rm MS}}^{N_f=3} = 260$ MeV \cite{Gockeler:2005rv},
 where we assume a degenerated $N_f=3$ case.
The renormalization point $\mu$ is assumed to be in a range $\mu= \pi T$--$3\pi T$.
Then, the Debye mass as a function $T$ in the leading-order (LO) thermal perturbation theory 
 is given by:
\be
\frac{m_D^{\rm LO}(T)}{T} = \sqrt{ 1 + \frac{N_f}{6} } \  g_{\rm 2l} (T)
,
\ee
where we have neglected the quark mass effects.
Dashed lines in Fig.~\ref{fig:2} (left) represent $m_D^{\rm LO}(T)$ for $N_f=3$
 for the above range of the renormalization point.
We find that the LO Debye mass does not reproduce the lattice data.

A formula in the next-to-leading-order (NLO) is also 
available from the resummed hard thermal loop calculation \cite{Rebhan:1993az}:
\be
\frac{m_D^{\rm NLO}}{T} = 
\sqrt{ 1 + \frac{N_f}{6} } \
g_{\rm 2l}(T) \left[ 1 + 
g_{\rm 2l}(T) \frac{3}{2 \pi} \sqrt{\frac{1}{1+ N_f/6}}
\left(
\ln \frac{2 m_D^{\rm LO}}{m_{\rm mag}} - \frac{1}{2}
\right)
+ o(g^2)
\right]
,
\label{eq:m_D_NLO}
\ee
where $m_{\rm mag}(T) = C_m g^2(T) T$  is the magnetic screening mass.
Since the factor $C_m$ cannot be determined in perturbation theory due to the infrared problem,
  we adopt $C_m \simeq 0.482$ calculated in 
a quenched lattice simulation \cite{Nakamura:2003pu} as a typical value.
Solid lines in Fig.~\ref{fig:2} (left) represent the NLO results for $N_f=3$. 
We find that 
the NLO Debye mass is approximately 50 \% larger than the LO Debye mass
 and  shows a better agreement with the lattice data.

Finally we study the flavor dependence of $m_D$.
Fig.~\ref{fig:2} (right) shows $m_D$ in lattice simulations of $N_f=2+1$ QCD (this work),
 that of $N_f=2$ QCD with an improved Wilson-quark action 
 at $m_{\pi}/m_{\rho} = 0.65$ \cite{WHOT-M}, and
 that of the quenched QCD ($N_f=0$) \cite{WHOT-U}.
Arrows on the horizontal axis indicate critical temperatures for
 $N_f=2$ ($T_c \sim 186$ MeV at $m_{\pi}/m_{\rho} = 0.65$ \cite{WHOT-FM}) and 
  $N_f=0$ ($T_c \sim 290$ MeV).
We find that $m_D$ for $N_f = 2+1$ is comparable
 to that for $N_f=2$, whereas it is larger than that for $N_f=0$.
%This implies that dynamical light flavors are important for $m_D$, 
A similar result was obtained with a staggered-type quark action \cite{Petrov:2007ug}.

%--------------------------------------------------------------------
\section{Summary}

We studied the free energy between a static quark and an antiquark 
at finite temperature in lattice QCD with $2+1$ flavors of improved Wilson quarks 
on $32^3 \times (12$--$4)$ lattices in the high temperature phase. 
We adopted 
the fixed scale approach which enables us to directly compare the free-energies at different
 temperatures without any adjustment of the overall constant. 
At short distances, the heavy-quark free energies, evaluated from the
 Polyakov-line correlations in the color-singlet channel,
 show universal Coulomb-like behavior common to that of the heavy-quark potential 
  at zero temperature evaluated from the Wilson-loop operator.
This is in accordance with the expected insensitivity of short distance physics to the temperature.
At long distances, the heavy-quark free energies approach
 to twice the single-quark free energies calculated from the thermal average
of a Polyakov loop. Also,
we  extracted the Debye screening mass $m_D(T)$ from the heavy-quark free energy
and found that the next-to-leading order  thermal perturbation theory is required 
to explain the magnitude of $m_D(T)$ on the lattice.
Comparison to the previous results at $N_f=2$ and $N_f=0$,
 shows that the dynamical light quarks have sizable effects on the value of $m_D(T)$.

\paragraph{Acknowledgments}
We thank the members of CP-PACS and JLQCD Collaborations
 for providing us with the data at zero temperature.
This work is partially supported 
by Grants-in-Aid of the Japanese Ministry
of Education, Culture, Sports, Science and Technology, 
(Nos. 17340066, 18540253, 19549001, 20105001, 20105003, 20340047, 21340049). 
SE is supported by U.S. Department of Energy
(DE-AC02-98CH10886). 
This work is supported 
also by the Large-Scale Numerical Simulation
Projects of CCS/ACCC, Univ. of Tsukuba, 
and by the Large Scale Simulation Program of High Energy
Accelerator Research Organization (KEK) 
Nos.08-10 and 09-18.

\end{document}